\documentclass[a4paper,11pt]{article}
\usepackage{epsfig}
\usepackage{amssymb}
\begin{document}
\thispagestyle{empty}

\newcommand{\etal}  {{\it{et al.}}}  
\def\Journal#1#2#3#4{{#1} {\bf #2}, #3 (#4)}
\def\PRD{Phys.\ Rev.\ D}
\def\NIMA{Nucl.\ Instrum.\ Methods A}
\def\PRL{Phys.\ Rev.\ Lett.\ }
\def\PLB{Phys.\ Lett.\ B}
\def\EPJ{Eur.\ Phys.\ J}
\def\IEEETNS{IEEE Trans.\ Nucl.\ Sci.\ }
\def\CPCD{Comput.\ Phys.\ Commun.\ }

\smallskip

\bigskip

{\LARGE
\begin{center}
Comments on gauge unparticles
\end{center}
}


\begin{center}
{\Large
 G.A. Kozlov

 }
\end{center}


\begin{center}
\noindent
 {
 Bogolyubov Laboratory of Theoretical Physics\\
 Joint Institute for Nuclear Research,\\
 Joliot Curie st., 6, Dubna, Moscow region, 141980 Russia\\

 }
\end{center}


\begin{abstract}

\noindent
A field model for a quark and an antiquark binding is described.
Quarks interact via a gauge unparticle ("ungluons").
The model is formulated in terms of
Lagrangian which features the source field $S(x)$ which becomes
a local pseudo-Goldstone field of conformal symmetry - the pseudodilaton mode
and from which the gauge non-primary unparticle field
is derived by $B_{\mu}(x) \sim\partial_{\mu} S(x)$. Because the conformal
sector is strongly coupled, the mode $S(x)$ may be one of new states
accessible at high energies.
We have carried out an analysis of the important quantity that enters
in the "ungluon" exchange pattern - the "ungluon" propagator.

\end{abstract}

\section{Introduction}

\bigskip

It is evidently that unparticle phenomenon [1] and its
phenomenology have been widely overlooked and discussed in the
literature (see, e.g., the recent papers in [2] and the references therein).

To develop a model for quark and an antiquark binding we follow the
Georgi's idea [1] that a nontrivial scale invariant sector of
scale dimension $d$ might manifest itself at low energy as a nonintegral
number $d$ of massless unparticles. One of the physical realization
of unparticle imagination is to look at unparticle as a limiting case in
which the  unparticle fields and their mass spectra are given
in the tower of infinite number of particles [3].

We assume that the fields
of the hidden sector undergo dimensional transmutation at scale
$\Lambda$ generating scale invariant unparticle field.
It means that $\Lambda$ defines a border energy scale where unparticle field(s) can
interact with the Standard Model (SM) ones.

It is worth recalling at this stage that the interaction of the hidden sector
given by the scale invariant unparticle operator $O_{U}$ with dimension $d$
and the SM operator $O_{SM}$ of dimension $n$ is
\begin{eqnarray}
\label{e1}
\left (\frac{\Lambda}{M_{U}}\right )^{d_{UV} + n -4}\,\frac{O_{U}\,O_{SM}}
{\Lambda^ {d + n -4}},
\end{eqnarray}
where $M_{U}$ is the mass of messenger in the ultra-violet (UV) hidden sector
of dimension $d_{UV}$ possessing the infra-red (IR) fixed point.
If $O_{U}$ is a vector field operator $O_{\mu}$ it could couple to the matter field(s)
and its exchange between particles in the SM could lead to additional effect of
interactions.

We investigate the effects on the conformal sector from the gauge sector, and
will show that this leads to surprising new bounds on unparticle physics.
The main attention is to the effective operator of the type
($\hat A = \gamma_{\mu}\,A^{\mu}$)
\begin{eqnarray}
\label{e2}
\frac{g^{\star}\,\bar\psi (x)\,\hat O(x)\,\psi (x)}{\Lambda^{d-1}}, \,\,\,
g^{\star} = g\,\left (\frac{\Lambda}{M_{U}}\right )^{d_{UV}},
\end{eqnarray}
where $g$ is dimensionless and $\psi (x)$ being  the prototype spin- 1/2
spinor (quark) field. We assume that $O_{\mu}(x)$ transforms like a vector
operator under the gauge transformations, and thus, the term with (\ref{e2})
gives an action which is invariant under this transformations. The interaction
given in (\ref{e2}) implies that the unparticle can be exchanged between massive spinor
particles, and this exchange creates a new force, which we call "ungluon" force
added to the standard gluon force.


We shall not consider the quantization of gauge unparticle field in the
standard conventional manner for following reasons:\\
- to be consistent with experiment where no such unparticle has been identified;\\
- to avoid the IR problems in perturbation theory.

We attribute no dynamical degrees of freedom to the gauge unparticle. Instead we
regard the unparticle field as the object of a direct interaction between quark and
antiquark. We shall investigate the effects of the scale invariant sector from the
gauge field sector, and we will show that this leads to new bounds on unparticle physics.

The paper is organized as follows. In section 2 we introduce the effective Lagrangian
of the model. The "ungluon" propagator function in four-dimensional space-time (4D)
is given in section 3. In the last section we conclude with some remarks.

\section{Scale invariance and (pseudo)dilaton mode}
We start by discussing a model of unparticle physics, which is different from
the previously suggested models.
The gauge-invariant operator $Q_{\alpha\beta} (1,2)$ for quark $1$ and antiquark $2$ as
$4\times 4$ matrix in Dirac spinor indexes $\alpha$ and $\beta$ is
\begin{eqnarray}
\label{e3}
Q_{\alpha\beta} (1,2) = -\frac{1}{N}\,\bar\psi_{\beta}(2)\,U(2,1)\,\psi_{\alpha}(1),
\end{eqnarray}
where the unparticle operator $U(2,1)$ is given by unparticle field $B_{\mu}(x)$:
\begin{eqnarray}
\label{e4}
U(2,1) = \exp\left\{-i\,g\int_{0}^{1} ds\,B_{\mu}(x_{s})\,\frac{dx^{\mu}_{s}}{ds}\right\}.
\end{eqnarray}
Here, the straight path integration is taken through $x_{s} = x_{1} + s\,r$, where $r = x_{2} - x_{1}$
is the relative distance between fermi-particles.

With respect to the scale invariance, the Noether theorem tells about the existence of
the corresponding conserved dilatation current $J^{\mu}_{dil}$, where
$\partial_{\mu}J^{\mu}_{dil}= \theta^{\mu}_{\mu}$ being the energy-momentum tensor. In
conformally invariant theories the last expression is equal to zero, however due to quantum effects
conformal invariance is broken [4]
\begin{eqnarray}
\label{e401}
\theta^{\mu}_{\mu} =\sum_{q:quarks} m_{q}\,\bar\psi\,\psi + \frac{\beta (g)}{2\,g}\,F^{a\,\mu\nu}\,
F^{a}_{\,\mu\nu}.
\end{eqnarray}
Here, the function $\beta (g) = \mu\,\partial g(\mu)/(\partial\mu)$ governs the behaviour
of the running coupling $g$ with the scale $\mu$.
In the case of $SU(3)$ gauge theory coupled to $n_{f}$ massless fermions in the fundamental
representation [5]
\begin{eqnarray}
\label{e402}
\beta (g)= \left (\beta_{0}\frac{g^{3}}{16\,\pi^{2}} + \beta_{1}\frac{g^{5}}{(16\,\pi^{2})^{2}}\right ),
\end{eqnarray}
where $\beta_{0} = -[11- (2/3)\,n_{f}]$, $\beta_{1} = -[102- (10+ 8/3)\,n_{f}]$. If $g$ is small
enough at high energy, it will increase as the renormalization scale $\mu$ decreases until the
fixed point $\beta (g) = 0$ is encountered at some $g = g^{\star}$. This is an IR fixed-point of the
renormalization group flow. Hence, at energy $E < \Lambda _{U} < M_{U}$ the effective theory becomes
scale-invariant.

It has been demonstrated [6] that
the coupling of gluodynamics to the conformal background gravity, described by a
single scalar field (dilaton),
leads to the fact  that theory is conformally invariant in any dimension.

In the Lagrangian framework we introduce two vector fields:
 $B_{\mu}(x)$ and $C_{\mu}(x)$, of which only
one - an unparticle gauge ("ungluon") $B_{\mu}(x)$ - field will interact directly with the
quark field $\psi (x)$ with mass $m$ ($C_{\mu}(x)$ being the auxiliary field).
The Lagrangian density is
\begin{eqnarray}
\label{e5} L  = -\frac{1}{4}\,B_{\mu\nu}\,B^{\mu\nu} +
\bar\psi (i\,\hat\partial - m - \frac{g^{\star}}{\Lambda^{d-1}}\,
\hat B )\psi - \frac{\alpha^{2}}{2\,\Lambda^{d-1}}\,B_{\mu\nu}\,
C^{\mu\nu} \cr
-\xi\,\frac{1}{\Lambda^{d-1}}\,\left (\partial_{\mu}\,B^{\mu}\right )\,
\left (\partial_{\nu}\,C^{\nu}\right ) - \frac{1}{2}\,\mu^{2}\,C_{\mu}\,C^{\mu} +
\frac{m^{2}_{S}}{2} \left (\frac{1}{\Lambda^{d-1}}\,B_{\mu} -\partial_{\mu} S\right )^{2},
\end{eqnarray}
where $B_{\mu\nu} = \partial_{\mu}\,B_{\nu} - \partial_{\nu}\,B_{\mu}$,
$C_{\mu\nu} = \partial_{\mu}\,C_{\nu} - \partial_{\nu}\,C_{\mu}$;
$\alpha$ is dimensionless; $\xi$ is the gauge
parameter; $\mu$ is a dimensional coupling constant - a mass parameter.
The scalar field $S(x)$ serves as the conformal compensator with mass $m_{S}$.
In real world the scale invariance is lost, particles possess finite mass and sizes.
It is therefore tempting to formulate an effective theory of broken scale invariance
also in terms of the corresponding pseudo-Goldstone boson of spontaneously broken
(approximate) scale invariance. For example, a light Higgs boson $h$ itself can be
identified with the pseudodilaton through the relation $S = \sqrt{h\,h^{+}}$.
Since the scale transformation does not affect any quantum numbers, the corresponding
particle can be, e.g., a scalar glueball or, perhaps, a $\sigma$- or $f_{0}$- mesons.

In this letter, we suppose that electroweak symmetry breaking is triggered by
a spontaneously breaking of scale symmetry (near conformal sector) at the energy scale
$f\geq v$ [7,8], where $v$ being the vacuum expectation value of the Higgs boson in the SM.
The spectrum of states at the electroweak (EW) scale $\Lambda_{EW}\sim 4\,\pi\,v$
contains a scalar particle (resonance),
the pseudo-Goldstone boson, pseudodilaton, of conformal symmetry breaking. This particle
is associated with the EW singlet scalar field $S(x)$, the dilaton mode.
The typical
pattern is provided by new strongly coupled, nearly conformal dynamics at a scale
of Conformal field theory (CFT) $\Lambda_{CFT}\sim 4\,\pi\,f$ which then flows into
EW sector at $\Lambda_{EW}$. The mass $m_{S}$ is naturally light, $m_{S}\sim\gamma\,f$,
where $\gamma$ is the parameter that controls deviations from exact scale invariance.
The dilaton becomes massless when conformal symmetry is recovered. Hence, the light scalar
resonance is a distinguishing feature of nearly conformal dynamics.

The prospects for distinguishing the dilaton mode from a minimal Higgs boson at the
LHC and ILC is presented in [8].

\section{"Ungluon" propagator}

>From the physical point of view, observable quantities may be defined as those
that are invariant under the gauge transformations of the second kind,
\begin{eqnarray}
\label{e6}
\psi \rightarrow\psi\,\exp(i\,g\,\lambda),\,\, B_{\mu}\rightarrow B_{\mu}
+ \Lambda^{d-1}\,\partial_{\mu}\lambda, \,\,\,C_{\mu}\rightarrow C_{\mu}
+ \partial_{\mu}\lambda,\,\,\,S\rightarrow S + \lambda
\end{eqnarray}
and $\lambda$ satisfies $\partial ^{2}\lambda = 0$.

The equations of motion read
\begin{eqnarray}
\label{e7}
\nabla^{2} B_{\mu} - \left (1- \frac{\xi}{\alpha}\right )\partial_{\mu}
(\partial\cdot B) =
 \frac{\mu^{2}}{\alpha}\frac{1}{\Lambda ^{1-d}}\,C_{\mu} ,
\end{eqnarray}
\begin{eqnarray}
\label{e8}
\nabla^{2} C_{\mu} - \left (1- \frac{\xi}{\alpha}\right )\partial_{\mu}
(\partial\cdot C)+ \frac{m^{2}_{S}}{\alpha}
\left (\frac{1}{\Lambda^{d-1}}B_{\mu} - \partial_{\mu}S\right )  = J^{\star}_{\mu} ,
\end{eqnarray}
\begin{eqnarray}
\label{e9}
B_{\mu} = \Lambda^{d-1}\,\partial_{\mu}S,
\end{eqnarray}
from which one can easily find the principal equations for fields $B_{\mu}$ and $S$
\begin{eqnarray}
\label{e10}
(\nabla^{2})^{2} B_{\mu} - \left (1- \frac{\xi^{2}}{\alpha^{2}}\right )\,\nabla^{2}
\partial_{\mu}(\partial\cdot B) =
\frac{\mu^{2}}{\alpha}\,\frac{1}{\Lambda ^{1-d}}\,J^{\star}_{\mu} ,
\end{eqnarray}
\begin{eqnarray}
\label{e101}
(\nabla^{2})^{2} S - \frac{\mu^{2}}{\xi}(\partial \cdot C) = 0 ,
\end{eqnarray}
where the current $J^{\star}_{\mu} = \alpha^{-1}\,g^{\star}\,\bar\psi\gamma_{\mu}\psi - \partial^{\nu}\,B_{\nu\mu}$ is
conserved.
Note that the field $C_{\mu}$ does not commute with $B_{\mu}$. If $\mu\rightarrow 0$, $S(x)$
becomes the dipole pseudodilaton field obeying the equation
\begin{eqnarray}
\label{e11}
(\nabla^{2})^{2} S (x) =0.
\end{eqnarray}
Few words concerning the higher derivative scalar field theory. We have already mentioned that
on the classical level conformal symmetry is manifested by the fact that the
stress-energy tensor is traceless. On the other hand, at quantum level the conformal symmetry
is broken which gives rise to conformal (trace) anomaly. It was found that for
4-dimensional higher derivative scalar field theory the trace anomaly can be obtained
from the action $E=\int d^{4} x \sqrt {-g}\,S\Delta ^{4}S$, where $\Delta^{4}
\equiv(\nabla^{2})^{2} - 2R^{\mu\nu}\nabla_{\mu}\nabla_{\nu} +(2/3)R\nabla^{2} -
(1/3) (\nabla_{\mu}R)\nabla^{\mu}$. Operator $\Delta^{4}$ contains the conformally
covariant structure for a fourth-order differential operator [9].
The field $S(x)$ in (\ref{e11}) provides an instructive and useful control over UV and
IR divergences for its free propagator. More detailed consideration of
dipole-field scalars are given in [10-14].

In the framework of  decomposition scheme [3], the unparticle interaction with
the SM quarks in (\ref{e5}) becomes
\begin{eqnarray}
\label{e12}
\frac{1}{\Lambda^{d-1}}\,g^{\star}\,\bar\psi\,\gamma_{\mu}\,\psi\sum_{k=1}^{k=\infty}\,f_{k}\,\partial^{\mu}\,S_{k},
\end{eqnarray}
where
\begin{eqnarray}
\label{e13}
f^{2}_{k} = \frac{A_{d}}{2\,\pi}\,\left
(m^{2}_{S_{k}}\right )^{d-2}\,\Delta^{2}_{s},\,\,\,
A_{d} = \frac{16\,\pi^{5/2}}{(2\,\pi)^{2d}}\,\frac{\Gamma (d+1/2)} {\Gamma
(d-1)\,\Gamma(2d)}.
\end{eqnarray}
The pseudodilaton fields $S_{k}$ are characterized by the mass squared
$m^{2}_{S_{k}} =k\,\Delta^{2}_{s}$ as
$\Delta_{s}\rightarrow 0$. Therefore the coupling of each $S_{k}$ to the SM quarks is
proportional to $\Delta_{s}$ and vanishes in the continuum limit $\Delta_{s}\rightarrow 0$.

The most general form of the commutator for free $B_{\mu}$ field is (we put $\xi =1$,
$\alpha =1$ so simplicity)
\begin{eqnarray}
\label{e14}
[B_{\mu}(x), B_{\nu} (y)] = i\,g_{\mu\nu}\left [-\mu^{2}\,\Lambda^{2(d-1)}\,E(x-y) +
c\,D(x-y) + const\right ],
\end{eqnarray}
where $c$ is an arbitrary real number; the commutator functions, namely, the invariant
function $E(x-y)$
and the Pauli-Jordan function $D(x-y)$ are [15,11]
\begin{eqnarray}
\label{e15}
E(x) = i\int 2\pi sign(p^{0})\delta^{\prime} (p^{2})
e^{-ipx}\frac{d^{4}p}{(2\pi)^{4}} = (8\,\pi)^{-1}\,sign (x^{0})\,\theta (x^{2}),
\end{eqnarray}
\begin{eqnarray}
\label{e16}
D(x) = \nabla^{2} E(x) = (2\,\pi)^{-1}\,sign (x^{0})\,\delta (x^{2})
\end{eqnarray}
with their properties
\begin{eqnarray}
\label{e17}
E(0,\vec x) = \partial_{0}E(x)_{\vert x^{0} =0} =
\partial^{2}_{0}E(x)_{\vert x^{0} =0}= 0, \,\,\partial^{3}_{\mu}E(x)_{\vert x^{0} =0}=
g_{\mu\,0}\delta^{3}(\vec x),
\end{eqnarray}
\begin{eqnarray}
\label{e18}
\nabla^{2} D(x)=0,\,\,D(0,\vec x) = 0,\,\,
\partial_{0}D(0,\vec x)=\delta^{3}(\vec x).
\end{eqnarray}
The form (\ref{e14}) ensures the equal-time canonical commutation relation (CCR)
\begin{eqnarray}
\label{e19}
[B_{\mu}(x), \pi_{B_{\nu}} (y)]_{x^{0}=y^{0}} = i\,g_{\mu\nu}\,\delta^{3}(\vec x - \vec y),
\end{eqnarray}
where
\begin{eqnarray}
\label{e20}
 \pi_{B_{\mu}} = \Lambda^{1-d}\,\left [\partial_{\mu} C_{0} - \partial_{0} C_{\mu} -
 g_{0\mu}\,(\partial\cdot C) + \partial_{\mu}B_{0} - \partial_{0}B_{\mu}\right ].
\end{eqnarray}
The next step is to decompose $E(x)$ into its negative
($E^{-}(x)$) - and positive ($E^{+}(x) = [E^{-}(x)]^{*} = -
E^{-}(-x) $) -frequency parts, each of which is analytic in the
past and future tubes: $E(x) = E^{-}(x) + E^{+}(x)$ with [15]
\begin{eqnarray}
\label{e21} E^{-}(x) = i\int 2\pi\theta (p^{0})\delta^{\prime}
(p^{2}) e^{-ipx}\frac{d^{4}p}{(2\pi)^{4}} =
-\frac{i}{(4\pi)^{2}}\,\ln \frac{l^{2}}{-x^{2} + i\epsilon x^{0}}
\cr = -\frac{i}{(4\pi)^{2}} [\ln\vert\kappa^{2} x^{2}\vert + i\pi
sign (x^{0})\theta (x^{2})].
\end{eqnarray}
Here, $l$ is an arbitrary length scale with dimension minus one in
mass units and introduced in the logarithmic function $\ln
[-(x^{0} - i\epsilon)^{2} + \vec x^{2} ]$ for dimensional reason
and $\kappa\sim l^{-1}$ being the mass parameter of the IR
regularization. The origin of $l$ becomes more transparent in
momentum space. Note that the distribution $\theta
(p^{0})\delta^{\prime} (p^{2})$ in (\ref{e21}) is well-defined
only with the basic functions $u(p)$ having the properties: $u(p)
= 0$ at $p = 0$.


The time-ordered two-point Wightman function (TPWF) for $B_{\mu}$ field is
\begin{eqnarray}
\label{e22} W_{\mu\nu}(x) = \langle 0\vert T\,B_{\mu}(x)\,B_{\nu}(0)\vert 0\rangle =
\theta (x^{0})\omega_{\mu\nu} (x) + \theta (-x^{0})\omega_{\mu\nu} (-x),
\end{eqnarray}
where TPWF $\omega_{\mu\nu}(x) = \langle 0\vert B_{\mu}(x)B_{\nu}(0)\vert 0\rangle$ is
\begin{eqnarray}
\label{e23}
\omega_{\mu\nu}(x) =
i\,g_{\mu\nu}\left [\frac{-\mu^{2}}{\Lambda^{2(1-d)}}\,E^{-}(x-y) +
c \,D^{-}(x-y) + const\right ]
\end{eqnarray}
with $D(x) = D^{-}(x) - D^{-}(-x)$.


Using results obtained in paper [14] we get the propagator function (\ref{e22}) in
4D momentum space in any local covariant gauge
\begin{eqnarray}
\label{e24}
\tilde W_{\mu\nu} (p) = \left [g_{\mu\nu} - \left (1-\frac{1}{\xi^{2}}\right )\,\frac{p_{\mu}\,p_{\nu}}
{p^{2}}\right ]\,\mu^{2}\,\tau (p;\kappa^{2}),
\end{eqnarray}
where
\begin{eqnarray}
\label{e25}
\tau (p;\kappa^{2}) = \sum _{k=1}^{k = \infty}\, \lim_{\lambda_{k}\rightarrow 0}\,f^{2}_{k}
\left [\frac{1}{(p^{2} - \lambda^{2}_{k} + i\epsilon)^{2}} + \frac{i}{(4\,\pi)^{2}}\,
\ln\frac{\lambda^{2}_{k}}{\kappa^{2}}\,\delta ^{4}(p)\right ].
\end{eqnarray}
The distribution $\lim_{\lambda^{2}_{k}\rightarrow 0} [1/(p^{2} -
\lambda^{2}_{k} + i\epsilon)^{2}]$ is defined only on a particular
subspace of the space of complex Schwartz test functions on $\Re
^{4}$, namely on those (test) functions $f(p)$ such that $f(0)
=0$. The set of its extensions onto the whole space is a
one-parameter set of functionals parameterized by $\kappa$.
It means that the set of Lorentz-invariant,
causal extensions of this distribution to those not vanishing at $p = 0$
constitute a new $\kappa$ - parameter  family.

In the continuum limit the final result for $\tau (p;\kappa^{2})$ is
\begin{eqnarray}
\label{e26}
\tau (p;\kappa^{2}) = \frac{A_{d}}{2\,\pi}\left \{\frac{(2-d)\pi}{(-1)^{d-1}}\,cosec [(d-1)\,\pi]\,
\frac{1}{(p^{2}+i\,\epsilon)^{3-d}} \right.\cr
\left. + \lim_{\epsilon\rightarrow 0}\frac{i}{(4\,\pi)^{2}}\,\frac{\ln\epsilon-
\gamma- \psi\left(\frac{3}{2}-d\right )}{\left(\frac{3}{2}-d\right )\epsilon^{3/2 -d}}\,
\frac{1}{(\kappa^{2})^{3/2 -d}}\,\delta^{3}(\vec p)\right \},
\end{eqnarray}
which is valid for $1< d < 3/2$; $\gamma = - \Gamma^{\prime}(1)\simeq 0.577$, $\psi (x) =
\Gamma^{\prime}(x)/\Gamma (x)$ is the digamma-function. This formula accompanied by (\ref{e24})
gives the main result for "ungluon" propagator which has the asymptotic behaviour in the form
$\sim g_{\mu\nu} /(p^{2})^{2-d}$ at $\xi =1$.

There are prices that must be paid for maintaining new result: i) the Fourier
transformation of TPWF $\omega_{\mu\nu} (x)$
contains $\delta^{\prime}(p^{2})$-function which is the consequence of non-unitarity character
of the model ($\delta^{\prime}$ is not a measure); ii) the spectral function
of the first term in expansion (\ref{e23}) gives an indefinite metric and hence
the translations become pseudounitarity (see paper [10] for details).


\section{Conclusions}

The gauge unparticle model has been studied, using canonical quantization in the
framework of decomposition scheme. One of the main objects is the pseudodilaton
field $S(x)$ which governs the "ungluon" field
$B_{\mu}(x) = \Lambda^{d-1}\,\partial_{\mu} S(x)$. We have found the propagator function
$\tilde W_{\mu\nu}(p)$ (\ref{e24}) of "ungluon" field in 4D momentum space in any local covariant gauge. The
dipole-type "ghost" behaviour of $\tilde W_{\mu\nu}(p)$ is evident. It is shown that no mass of
"ungluon" field appeared in (\ref{e26}).

The non-unitarity character of the model is the direct consequence of i) the form
of the Lagrangian density (\ref{e5}), ii) scale and gauge invariance
of the model, iii) spontaneous breaking of scale invariance, iv) the form of the equation of
motion and related commutator for the field $B_{\mu} (x)$ (\ref{e14})
which ensures the CCR (\ref{e19}).

It follows from our investigation that the degrees of freedom called "unparticles" in [1] do,
indeed, in the gauge sector very much behave like the "ghost" fields.

It would be especially interesting to see how the inclusion of the correction to
non-perturbative potential arising from exchange of "ungluons" (at least to lowest order)
affects QCD physics. This item is left for future paper.

\section{Acknowledgments}

It is a pleasure to thank Fermilab  I visited during the
course of this work.


\end{document}